\documentstyle[aps]{revtex}
\begin{document}
\title{Analytical expressions for the local-field factor $G({{q}})$ and
the exchange-correlation kernel $K_{xc}({r})$ of the homogeneous electron gas.}

\author{Massimiliano Corradini, Rodolfo Del Sole, Giovanni Onida, 
and Maurizia Palummo}
\address{
Istituto Nazionale per la Fisica della Materia, Dipartimento di Fisica
   dell'Universit\`a \\ di Roma ``Tor Vergata'',
   Via della Ricerca Scientifica, I--00133 Roma,
   Italy }

\protect{\bigskip}
\bigskip
\draft
\maketitle
\bigskip
\begin{abstract}
We present an analytical expression for the local field factor 
$G({q})$ of the homogeneous electron gas which reproduces recently
published Quantum Monte--Carlo data by S. Moroni, 
D.M. Ceperley, and G. Senatore 
[Phys. Rev. Lett. {\bf{75}}, 689 (1995)],
reflects the theoretically known 
asymptotic behaviours for both small and large ${{q}}$ limits, and allows to
express the exchange-correlation kernel $K_{xc}$ analytically in
both the direct and reciprocal spaces. The last property is
particularly useful in numerical applications to real solids.
\end{abstract}

\narrowtext
The static local field factor $G({q})$ of the homogeneous electron gas 
(HEG) is an important quantity  because it represents the extent to which the
particle interactions affect the exchange and correlation 
properties of this idealized system.
The importance of $G({{q}})$ in calculating the properties
of real materials stems from the fact that it can be used as a
key input in density functional calculations \cite{kohn,delsole}. 
In fact, approximations of the unknown exchange-correlation
energy functional of real (inhomogeneous) systems involve the
second functional derivative of $E_{xc}[n]$, or exchange correlation kernel
\begin{equation}
K_{xc}(n_{o};|{\bf{r}}-{\bf{{r}^{'}}}|)=\frac{\delta^{2} E_{xc}[n] }
{\delta n({\bf{r}})\delta n({\bf{r}^{'}})}|_{n_{o}},
\end{equation}
where $n_{o}$ is the HEG density.
On the other hand, the local field factor and the exchange correlation 
kernel are simply related in Fourier space by \cite{tosi}:
\begin{eqnarray}\label{eq_kvg}
K_{xc}^{FT}({{q}})&\equiv&\int d^{3}{\bf{r}} 
e^{-i\bf{q} \cdot \bf{r}}K_{xc}({r})\\ \nonumber 
&=&-v_{c}({{q}})G({{q}}),
\end{eqnarray}
where $v_{c}({{q}})=4\pi e^{2} / q^{2}$ is  the Coulomb potential.

Various approaches to obtain expressions of
the local field factor were investigated in the 
past \cite{moroni,ichi,farid,bretonnet,schulke,lee,hong}. 
Moroni $et$ $al.$ \cite{moroni} obtained, by quantum Monte
Carlo (QMC) simulation, the zero temperature local field factor $G({{q}})$ of 
the homogeneous electron gas at densities $n_{o}$
corresponding to $r_{s}=2, 5,10$ 
($n_{o}=3/(4 \pi a_{B}^{3}r_{s}^{3})$, where  $a_{B}$ is the Bohr radius). 
They also
gave an analytical formula for $G({{q}})$ which fits the 
QMC computed value, and has the right asymptotic limits at small and large
wave
vector, $i.e.$:
\begin{equation}\label{eq_smallq}
G({{q}}) \sim AQ^{2} \hspace{2cm} for \hspace{.2cm}{{q}}\rightarrow 0,
\end{equation}
where $Q=q/k_{F}$, $k_{F}$ is the Fermi wave vector,
\begin{equation}
A=\frac{1}{4}-\frac{k_{F}^{2}}{4\pi e^2}\frac{d \mu_{c}}{d n_{o}}, 
\end{equation} 
$\mu_{c}$ being the correlation contribution to the chemical 
potential; and
\begin{equation}\label{eq_largeq}
G({{q}}) \sim CQ^{2}+B \hspace{2cm} for \hspace{.2cm}{{q}}\rightarrow \infty,
\end{equation} 
where 
\begin{equation}
C=\frac{\pi}{2 e^{2} k_{F}}\frac{-d (r_{s}\epsilon_{c})}{d r_{s}}, 
\end{equation}
and
$\epsilon_{c}$ is the correlation energy per particle. In our
calculations we use the parametrization of $B$ from 
Ref.\cite{moroni}:
\begin{equation} 
B(r_{s})=\frac{(1+a_{1}x+a_{2}x^{3})}{(3+b_{1}x+b_{2}x^{3})},
\end {equation}
 where 
$x=r_{s}^{1/2}$, $a_{1}=2.15$, $a_{2}=0.435$, $b_{1}=1.57$, and 
$b_{2}=0.409$, valid for $r_{s}$ in the range 2--10\cite{moroni2}.

The $Q^{2}$ behaviour of the local-field factor at large $Q$ has been
overlooked for a long time. It has been demonstrated in Ref. \cite{holas},
and a clear discussion of its origin is presented 
in Ref.\cite{moroni}: its coefficient $C$ is related 
to the change in kinetic energy in going from non-interacting 
(Kohn-Sham) electrons to interacting (real) electrons.

In this paper we fit the values of $G({{q}})$ computed in Ref.\cite{moroni}
in such a way to obtain a simple analytical expression of 
both $K_{xc}^{FT}({{q}})$ and $K_{xc}({{r}})$, where 
$K_{xc}({{r}})$ is the
exchange correlation kernel in real space. 
Our formula for $G({{q}})$ is based on Lorentzian and Gaussian functions,
and reads:
\begin{equation}\label{eq_g}
G({{q}})=CQ^{2}+\frac{BQ^{2}}{g+Q^{2}}+\alpha Q^{4}e^{-\beta Q^{2}},
 \end{equation}
where $g=B/(A-C)$ and the two parameters $\alpha$ and 
$\beta$ are fitted in order to minimize the differences with the
numerical results of Ref.\cite{moroni}. In particular, the best results
are obtained taking: 
\begin{equation}
\alpha=\frac{1.5}{r_{s}^{\frac{1}{4}}}\frac{A}{Bg},
\end{equation}
\begin{equation}
\beta=\frac{1.2}{Bg}.
 \end{equation}
Note that the Lorentzian contribution in Eq.\ref{eq_g} is a simple 
Hubbard-like
term. This term alone yields already $qualitative$ agreement with 
the numerical data of Ref.\cite{moroni}.
Adding the Gaussian term allows us to reproduce $quantitatively$ the
numerical evaluation of Ref.\cite{moroni}.
In panels $a$, $b$, and $c$ of Fig.1, $G(q)$ given by Eq.\ref{eq_g} is compared with the 
QMC results of Ref.\cite{moroni} for the unpolarized electron gas. 
The agreement is satisfactory, and is
globally of the same quality as that obtained with the interpolation
formula originally proposed in Ref. \cite{moroni}. In panel $d$ of 
Fig.1, we extend the comparison to QMC data for the fully spin polarized 
electron gas at $r_{s}=100$ \cite{private}. Despite the fact that this 
value is well beyond the range considered in Ref. \cite{moroni} for
the parametrization of B (Eq.7), and that in Eq.8 we neglect spin
polarization effects, 
the agreement is still fairly good.

Using Eq.\ref{eq_kvg}, and after Fourier transforming, we obtain the
expression of the exchange correlation kernel $K_{xc}(r)$ in real space:   
\begin{equation}\label{eq_r}
K_{xc}({{r}})=-\frac{4\pi e^{2} C}{k_{F}^2 }\delta^{3}({{r}})+
\frac{\alpha k_{F}}{4 \pi^2 \beta}
(\frac{\pi}{\beta})^{\frac{3}{2}}[\frac{k_{F}^2 r^2}{2 \beta}-3]
e^{-\frac{k_{F}^2 r^2}{4\beta}}-B\frac{e^{-\sqrt{g} k_{F} {r}}}{{r}}.
\end{equation}
In Fig.2, we compare this form of $K_{xc}$ (without the first term,
which contains a three-dimensional delta function) with the (numerical) 
FT of the kernel derived from the Utsumi and Ichimaru parametrization of
$G({{q}})$ \cite{ichi}. Besides the very desirable property of allowing
passage from real to reciprocal space and vice--versa without
numerical transforms, the present form has another advantage, which is
particularly useful in calculations for real solids: the absence of
long-range oscillations, at variance with the case of those
$G(q)$ which contain a logarithmic singularity for $q=2k_{F}$, as the
Ichimaru-Utsumi one (see Fig.2). 
This singularity is a peculiar property of the homogenous electron 
gas, originating from its spherical Fermi surface and from the 
absence of a gap between filled and empty states. Even in the HEG, its
existence is not certain, and it is likely not 
present in real materials \cite{nota1}. Hence it is better, given the present 
level of ignorance about the exchange-correlations kernel of real 
materials and the computational difficulties arising from such
a singularity, to use expressions of $G(q)$ which do not 
contain it, and consequently do not yield the long range oscillations of 
$K_{xc}({{r}})$.  

In conclusion, we have presented a new parametrization of published QMC
results for the local field factor $G(q)$ of the homogeneous electron
gas. Our analytical form fits the numerical data with the same
accuracy as the form originally proposed in Ref.\cite{moroni}, has the
right limiting behaviours for large and small $q$, and has the additional
advantage of being analytically Fourier-transformable, a property which
greatly simplifies its use in density-functional calculations for real
materials.

We are grateful to Saverio Moroni and Gaetano Senatore for
providing us their unpublished QMC data for $r_{s}=100$, and for
a critical reading of the manuscript.

\narrowtext
\newpage
\narrowtext
\begin{figure}
\caption {Panels $a$, $b$, and $c$: local field factor $G(Q)$, ($Q=q/k_{F}$) as computed 
according to Eq.
{\protect \ref{eq_g}},
in comparison with the QMC data for the unpolarized HEG of Ref. {\protect \cite{moroni}}.
Panel $d$: Eq.8 compared with QMC data for the fully spin polarized 
HEG {\protect \cite{private}}.}
\end{figure}

\begin{figure}
\caption {The exchange-correlation kernel $K_{xc}$ of the homogeneous
 electron gas as parametrized in the present work (full line),
 compared with the $K_{xc}$ of 
 Ref.{\protect \cite{ichi}} (dotted line), in 
 atomic units, plotted in reciprocal  and direct space (upper and lower panel,
respectively, with $Q=q/k_{F}$ and $R=r k_{F}$)}
\end{figure}


\bigskip

\newpage

\begin{references}

\bibitem{kohn}
P. Hoenberg and W. Kohn, Phys. Rev. {\bf{136}}, B864 (1964); W. Kohn and
L. J. Sham, Phys. Rev. {\bf{140}}, 1133 (1965).

\bibitem{delsole}
M. Palummo, G. Onida, R. Del Sole, and L. Reining, "Electronic structure
calculations beyond the local-density approximation: Application to
Silicon", Proc. 23$^{rd}$ Int. Conf. on the Physics of Semiconductors, Berlin,
World Scientific (1996), page 609.


\bibitem{tosi}
See, for instance, K. S. Singwi and M. P. Tosi, $ Solid$ $State$ $Physics$, 
Vol. 36,
edited by H. Ehrenreich, F. Seitz, and D. Turnbull (Academic, New York,
1981).



\bibitem{moroni}
S. Moroni, D.M. Ceperley, and G. Senatore, Phys. Rev. Lett. {\bf 75}, 689 (1995
).


\bibitem{ichi}
S. Ichimaru and K. Utsumi, Phys. Rev. B {\bf{24}}, 7385 (1981).

\bibitem{farid}
B. Farid, V. Heine, G. E. Engel, and I. J. Robertson,  
Phys. Rev. B {\bf{48}}, 11602 (1993). 

\bibitem{bretonnet}
J. L. Bretonnet and M. Boulahbak, Phys. Rev. B {\bf{53}}, 6859 (1996).

\bibitem{schulke}
W. Sch\"{u}lke, K. H\"{o}ppner, and A. Kaprolat,
Phys. Rev. B {\bf{54}}, 17464 (1996).

\bibitem{lee}
Keun-Ho Lee and K. J. Chang Phys. Rev. B {\bf{54}}, R8285 (1996).

\bibitem{hong}
J. Hong and Y. Shim, J. Phys. : Condens. Matter {\bf{5}}, 3431 (1993).

\bibitem{moroni2} G. Senatore, S. Moroni, D.M. Ceperley, {\it The
 Local Field of the Electron Gas}, in {\sl Physics of Strongly coupled
 plasmas}, ed. by W. D. Kraeft and M. Schlanges (World Scientific, 
Singapore, 1996) p. 429-434.

\bibitem{holas}
A. Holas, in {\it{Strongly Coupled Plasma Physics}}, edited by F. J. 
Rogers and H. E. DeWitt (Plenum, New York, 1987), p. 463.

\bibitem{private}
S. Moroni, D.M. Ceperley, and G. Senatore, private communications. At 
these small densities, the effects of spin-polarization are expected 
to be small.

\bibitem{nota1}
A plateau or weak minimum at about $k=2.5 k_{F}$, which could be related to 
such a singularity, is present only in the QMC data for the spin polarized 
HEG at $r_{s}=100$ (Fig. 1 $d$).

\end{references}
\end {document}
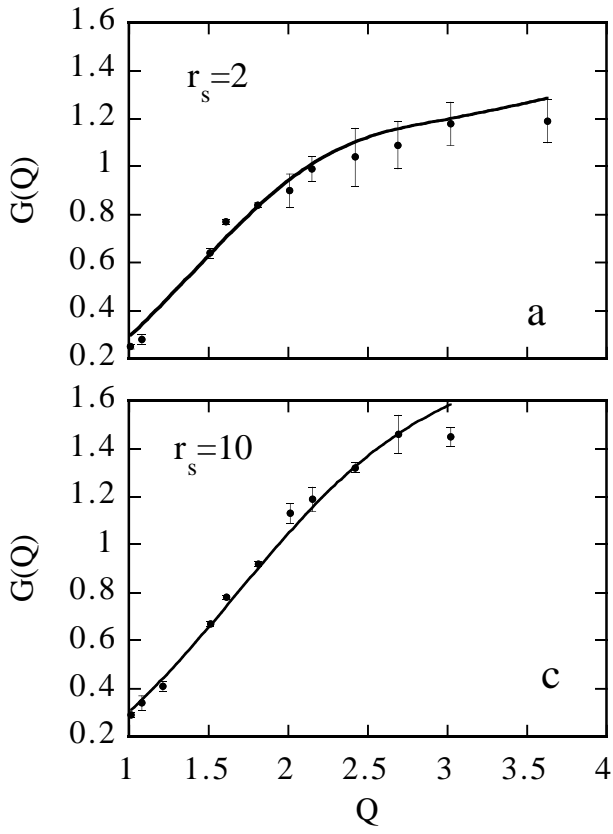
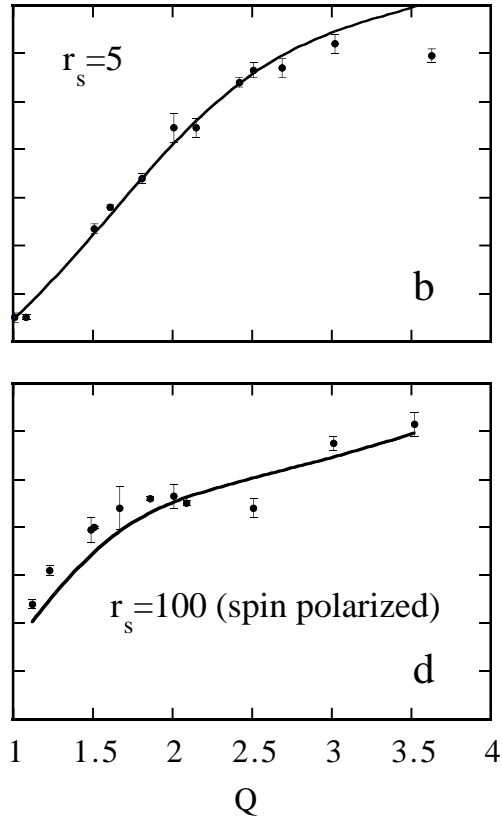
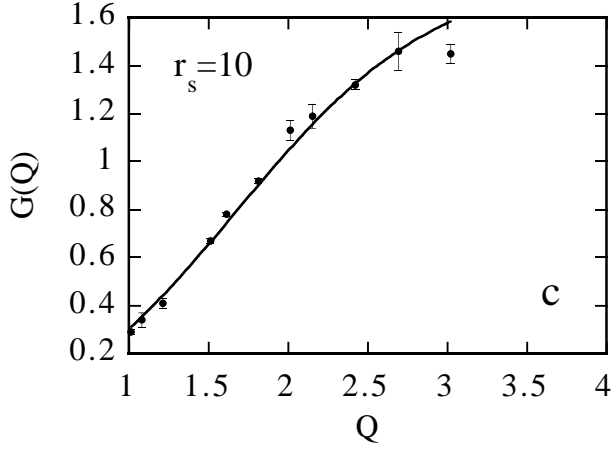
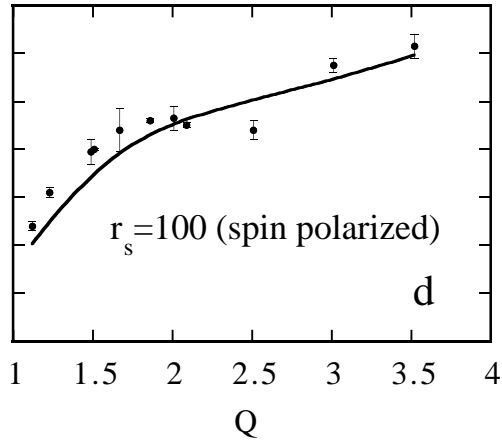
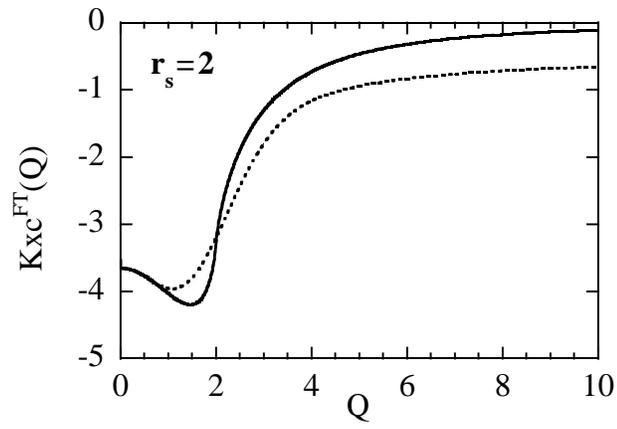
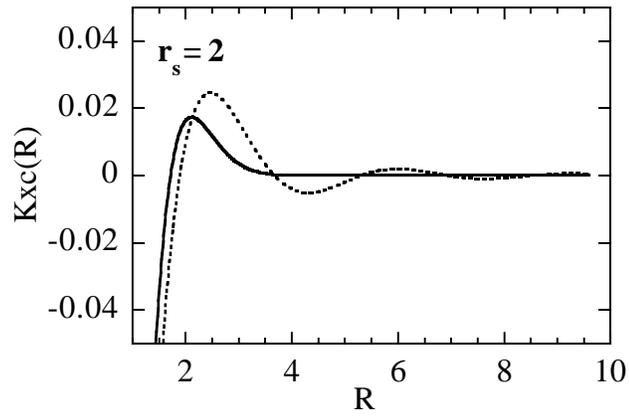
